\documentclass[twocolumn]{aastex631}
\usepackage{amsmath}
\usepackage[pagewise,displaymath, mathlines]{lineno}
\providecommand{\sorthelp}[1]{}

\begin{document}

\title{\boldmath Dynamical friction by coupled dark energy}

\author[0009-0008-6431-3002]{Nasrin Nari}
\author[0000-0002-9748-2928]{Mahmood Roshan}
\email{mroshan@um.ac.ir}
\affiliation{Department of Physics, Faculty of Science, Ferdowsi University of Mashhad, P.O. Box 1436, Mashhad, Iran}

\begin{abstract}
In this paper, we examine dynamical friction at galactic scales within the framework of coupled dark energy. This model posits dark energy as coupled quintessence, which maintains a minimal coupling to gravity but interacts non-minimally with both dark matter and baryonic matter. Since our focus is primarily on the Newtonian regime within galaxies, we begin by deriving the Newtonian limit of the model. Subsequently, we calculate the dynamical friction force using three different approaches. We demonstrate that, in the absence of interaction between dark energy and matter, standard quintessence does not generate any dynamical friction at the galactic scale. However, the presence of interaction does cause dynamical friction. By applying the resulting analytic expressions to a real self-gravitating system, namely the Fornax galaxy, and by implementing the constraints on the free parameter of the model obtained from galactic observations, we demonstrate that the coupled dark energy model leads to significant deviations from the standard cold dark matter model at galactic scales. On the other hand, if the cosmological constraints are assumed for the free parameter, the effects of the model are expected to be negligible at the galactic level, at least in dynamical friction.
\end{abstract}

\section{Introduction}\label{sec:intro}
Numerous alternative theories of gravity incorporate scalar fields to describe dark energy or dark matter. In the context of dark energy, notable examples include the standard quintessence model, coupled quintessence, K-essence, and chameleon scalar field theories \citep{amendola2010dark}. Conversely, for dark matter, one may refer to Tensor-Vector-Scalar (TeVeS) theory \citep{Bekenstein:2004ne}, MOdified Gravity (MOG) \citep{Moffat:2005si}, and the new relativistic theory of Modified Newtonian Dynamics (MOND) \citep{Skordis:2020eui}.

The primary objective of this paper is to examine dynamical friction within the specific coupled dark energy (CDE) model proposed by Amendola \citep{Amendola:1999er,Amendola:1999qq} in which the existence of a scalar field plays a key role. This model postulates a specific form of interaction between dark energy and matter. The interaction between dark energy and dark matter, resulting in an energy flux between the two components, has significant implications for cosmological evolution. This interaction can influence the duration of cosmological epochs, as demonstrated by \cite{Amendola:2003wa,Guo:2004xx,Wei:2007ws,Wei:2008nc,Cai:2009ht,Bolotin:2013jpa,Li:2014eha,Wang:2016lxa,Barros:2018efl,Gomez-Valent:2020mqn,Roy:2023uhc}. While a series of studies , \cite{Maccio:2003yk,Pettorino:2012ts,Pettorino:2013oxa,Xia:2013nua,Planck:2015bue,Costa:2016tpb,Liu:2023mwx,Hoerning:2023hks}, have demonstrated that observational constraints limit the viability of dark energy interaction models, recent investigations, including \cite{Yang:2022csz,Aboubrahim:2024spa}, have shown that such interactions can potentially alleviate some observational tensions within the standard cosmological model, such as the Hubble tension and the $S_8$ discrepancy. For a recent reviews on this model, we refer the reader to \cite{vanderWesthuizen:2023hcl} and \cite{Wang:2024vmw}.

The interaction between dark energy and dark matter can, in theory, influence the long-term internal evolution of galaxies. However, despite extensive research on this model in the context of cosmology, there has been comparatively less focus on its implications at galactic scales. One of the primary consequences of the presence of dark matter particles in galaxies is the emergence of dynamical friction, which can significantly impact the dynamics of stellar bars in spiral galaxies or globular clusters within dwarf galaxies. The concept of dynamical friction in self-gravitating systems was first developed by Chandrasekhar in \cite{Chandrasekhar:1943ys}. It has proven to be an invaluable tool for explaining various phenomena observed in galaxies. For example, it accounts for the presence of massive black holes in the central regions of galaxies, the absorption of satellite galaxies by their host galaxies, and the formation of binary black holes during galactic mergers, among other phenomena; see \cite{galacticdynamicsbook}. Another significant example is the exchange of angular momentum between the dark matter halo and the stellar bars in disk galaxies through the dynamical friction \citep{weinberg}. As a result, the pattern speed of the bar decreases over time \citep{sellwood}. Consequently, cosmological simulations predict ``slow'' bars at redshift $z=0$. In contrast, most observed bars are classified as ``fast''. This discrepancy presents a challenge for the standard model of cosmology \citep{Roshan:2021liy}.

The significance of dynamical friction in self-gravitating systems containing substantial amounts of dark matter has been investigated across various models of dark matter particles. Two notable cases are worth mentioning. First, \cite{Berezhiani:2019pzd} explored dynamical friction in the context of dark matter as a superfluid. In contrast to classical theories, this model demonstrates that dynamical friction persists even during subsonic motion within superfluid dark matter. In the same model, the role of dynamical friction in the evolution of a black hole binary has been investigated in \cite{Berezhiani:2023vlo}. Second, the fuzzy dark matter (FDM) is another model that has garnered considerable attention in the past seven years \citep{Hui:2016ltb}. In this framework, dark matter is composed of ultralight bosons that can induce quantum effects on a galactic scale. These quantum effects may reduce dynamical friction, thereby extending the orbital decay timescale of globular clusters in the Fornax dwarf spheroidal galaxy. This behavior helps to alleviate a long-standing puzzle regarding this galaxy \citep{Hui:2016ltb}. A more comprehensive study of dynamical friction in FDM has established a constraint on the mass of ultralight bosons needed to resolve the Fornax puzzle \citep{Lancaster:2019mde,Buehler:2022tmr}. On the other hand, \cite{Glennon:2023gfm} demonstrates that repulsive interactions between ultralight background particles can significantly reduce dynamical friction. This occurs because such interactions hinder the formation of a strengthened wake behind the moving body. Third, there are a few papers on dynamical friction in the context of self-interacting dark matter (SIDM) model as well. This model was first proposed by \cite{Spergel:1999mh} as a potential solution to discrepancies between CDM simulations and observational data. Recent investigations, including the analytical study in \cite{Alonso-Alvarez:2024gdz} and N-body simulations in \cite{Fischer:2024dte}, demonstrate that dark matter self-interactions have a substantial impact on the dynamical friction experienced by merging black holes. This impact is particularly significant in the final stages of the merger process, where it can influence the propagation of gravitational waves. Furthermore, these studies suggest that the inclusion of dark matter self-interactions may offer a potential solution to the ``final parsec problem''.

In the context of modified gravity theories that deny the existence of dark matter particles, dynamical friction serves as a crucial tool for distinguishing between modified gravity and cold dark matter. For example, \cite{Roshan:2021ljs} investigate dynamical friction within a specific theory of nonlocal gravity (NLG). In this framework, nonlocal gravitational effects replace the need for dark matter \citep{Hehl:2008eu}. The study demonstrates that dynamical friction in spiral galaxies is weaker compared to the standard cold dark matter model. This finding aligns with high-resolution N-body simulations conducted within the NLG framework \citep{Roshan:2021mfc}. The simulations in \cite{Roshan:2021mfc} approves a similar behavior in MOG. 

On the other hand, there are several papers investigating dynamical friction in MOND. Let's mention some of them: early studies , \cite{Ciotti:2004wb,Sanchez-Salcedo:2006tcj,Nipoti:2008fg}, demonstrate that dynamical friction in MOND can be greater than in Newtonian theory at least in the deep MOND regime, and that the relaxation time can be significantly shorter compared to Newtonian dynamics. Similar results have been reported by a more recent research \citep{DiCintio:2024kpr}. However, outside the deep MOND regime the situation is different and the friction force can be substantially reduced compared to Newtonian gravity. For example see \cite{Tiret:2007fy,Tiret:2007dd} where the merging time scale of Milky Way-type galaxies is explored. As another example, the galactic bar developed in the MONDian model in \cite{Roshan:2021liy} experiences much less dynamical friction compared to the dark matter case.

\cite{MichalBilek:2021} investigates conditions under which globular clusters maintain their orbits without forming a central nucleus. The analysis reveals that, for a single globular cluster in the MOND regime, similar to the Newtonian case, a core stalling phenomenon is observed. This stalling nucleus prevents the formation of a central nucleus in ultra diffuse galaxies (UDGs) . Furthermore \cite{Bilek:2024gva} examines the motion of globular clusters under the influence of dynamical friction in the presence of supernova feedback in MOND.

Now, let us return to the primary focus of this paper. The CDE model emerges as a hybrid model that incorporates both modifications to gravity and the presence of dark matter particles. In this framework, the scalar field primarily serves as dark energy. However, its coupling to matter (particularly to dark matter) may affect the internal dynamics of galaxies. This represents a distinctive aspect in which dark energy could have a significant role within galaxies. One of the prime locations to observe this intriguing feature is by exploring the impact of dynamical friction. To achieve this objective, we derive analytical expressions for the dynamical friction experienced by a massive point mass moving through both a medium composed of a sea of less massive particles and a gaseous medium. For the former case, we implement two different methods: the standard approach by Chandrasekhar \citep{Chandrasekhar:1943ys} and the gravitational wake approach developed in \cite{Tremaine:1984}. These approaches do not lead to precisely the same results because the latter proves to be more powerful and relaxes some restrictive assumptions of the earlier one. For the gaseous system, we adopt a fluid treatment and utilize the linearized hydrodynamical equations along with a modified version of the Poisson equation within the CDE model. In this approach, the Jeans analysis aids in determining the dynamical friction force; see \cite{Berezhiani:2019pzd} and \cite{Lancaster:2019mde}.

The outline of this paper is as follows: we begin by presenting the field equations of the CDE model in Sec. \ref{FE}. Sec. \ref{WFE} is dedicated to deriving the weak field limit of the model. Subsequently, in Sec. \ref{DF}, we derive the dynamical friction force for both particle and gaseous media. Finally, in Sec. \ref{APP}, we apply the results to the Fornax galaxy to investigate whether the CDE model can address the Fornax puzzle.

\section{Filed equations of CDE}\label{FE}
In Brans-Dicke theory the scalar field couples to the metric and is sourced by  matter. More importantly, the scalar field does not couple directly to the matter Lagrangian. However as shown in \cite{Damour:1990tw}, by applying a conformal transformation $ g_{\mu \nu}= e^{- 2 \kappa \beta \phi} ~ \overset{\sim}{g}_{\mu \nu}$ where $g_{\mu \nu}$ and $\overset{\sim}{g}_{\mu \nu}$ indicate the metric in the Einstein and Jordan frame respectively, $\kappa^2=8\pi G $, and $\beta $ is a parameter that determines the strength of coupling between matter and the scalar field, we can move to the Einstein frame. This transformation results in the  Brans-Dicke action taking the form
\begin{equation}\label{action}
	S=\int d^4x \sqrt{-g}\Big(\frac{c^4}{2 \kappa ^2} R + \mathcal{L}_\phi + \mathcal{L}_m\Big)
\end{equation}
where $g$ is the determinant of the metric, $R$ is the Ricci scalar, $\mathcal{L}_m$ represent the matter Lagrangian density and the scalar field Lagrangian density is
\begin{equation}
	\mathcal{L}_\phi=-\frac{1}{2} \nabla_\mu \phi \nabla^\mu \phi -V(\phi) .
\end{equation}
$V(\phi)$ is the potential function of the scalar field that is arbitrary. Hereafter we restrict ourselves to the Einstein frame and refer the reader to \cite{Damour:1990tw} for more details on the frame transformation. In the Einstein frame the energy-momentum tensor is no longer conserved. More specifically, the conservation equation of the energy-momentum tensors read
\begin{equation}
	\nabla_\nu \overset{(\phi)}{T^{\nu}_{~~\mu}}=-\kappa \beta_b \rho_b \nabla_\mu \phi-\kappa \beta_c \rho_c \nabla_\mu \phi
	\label{ceosf}
\end{equation}
\begin{equation}
	\nabla_\nu \overset{(c)}{T^{\nu}_{~~\mu}}=\kappa \beta_c \rho_c \nabla_\mu \phi
	\label{ceodm}
\end{equation}
\begin{equation}
	\nabla_\nu \overset{(b)}{T^{\nu}_{~~\mu}}= \kappa \beta_b \rho_b \nabla_\mu \phi
	\label{ceobm}
\end{equation}
where $\rho$ is energy density, and $c$ and $b$ subscripts denote cold dark matter and baryonic matter, respectively. The scalar field is postulated to play the role of dark energy. Therefore, the conservation equations simply imply that there is interaction between the two main ingredients of the cosmic matter-energy budget, namely dark energy and dark matter. The coupling parameters $\beta_c$ and $\beta_b$ regulate the strength of the interaction between matter and dark energy. In general, these parameters can depend on the scalar field; however, in this paper, we treat them as constant. By setting them to zero, we recover the standard uncoupled quintessence model. The case $\beta_c\neq \beta_b$ means that the fifth force is felt differently by matter and dark matter. Local gravitational experiments constrain $|\beta_b|$ to be small \citep{Amendola:1999er}. Conversely, the rotation curve data of 40 galaxies in the SPARC catalogue suggests that $\beta_b \beta_c=0.17\pm 0.04$ \citep{deAlmeida:2018kwq}. We will employ this constraint in our dynamical friction calculations, as we focus on the galactic scale. Additionally, there are other restrictive constraints on $\beta_c$ from cosmology (see, for example, \cite{Gomez-Valent:2020mqn,Pettorino:2013oxa,Xia:2013nua,Planck:2015bue}), which we will discuss in terms of their implications for our analysis in the Application section.

The metric field equation is
\begin{equation}
	\begin{split}
		&R_{\mu \nu}-\frac{1}{2} g_{\mu \nu}R=\frac{\kappa^2}{c^4}\Big(\overset{(c)}{T}_{\mu \nu}+\overset{(m)}{T}_{\mu \nu}+\overset{(\phi)}{T}_{\mu \nu}\Big)
	\end{split}
	\label{EnEq}
\end{equation}
where 
\begin{equation}
	\overset{(\phi)}{T}_{\mu \nu}=g_{\mu \nu}V(\phi)-\nabla_{\mu}\phi \nabla_\nu \phi+\frac{1}{2}g_{\mu \nu} \nabla_\gamma \phi \nabla^\gamma \phi
	\label{emtsf}
\end{equation}
and the energy-momentum tensor of the matter/dark matter is
\begin{equation}
	T_{\mu\nu}=\Big(\rho+\frac{p}{c^2}\Big)u_\mu u_\nu +p~g_{\mu\nu}
	\label{emt}
\end{equation}
where $\rho$ can be the energy density of baryonic or dark matter, $p$ is the corresponding pressure, and $u^{\mu}$ is the four-velocity. Notice that we assume that both dark matter and baryonic matter are described by perfect fluid.
On the other hand, by taking the variation of the action \eqref{action} with respect to the scalar field, we find the following field equation
\begin{equation}\label{feq1}
	\square \phi-V'(\phi)=-\kappa (\beta_b \rho_b+\beta_c \rho_c)
\end{equation}

\section{The Weak field limit of CDE }\label{WFE}
Although CDE serves as a cosmological model, our research aims to investigate its implications on the galactic scale. Therefore, it is imperative to derive the weak field limit of the model. Before proceeding, let's briefly review the weak field limit of GR in the presence of the cosmological constant $\Lambda$ playing the role of dark energy. In this case, the metric field equation is
\begin{equation}
	G_{\mu\nu}+\Lambda g_{\mu\nu} = \frac{\kappa^2}{c^4} T_{\mu\nu}
	\label{lambdafe}
\end{equation}
Deriving the Newtonian limit of equation \eqref{lambdafe} can provide useful insights for conducting analogous calculations in the CDE model, given that the latter constitutes a cosmological model under investigation with a focus on its galactic consequences.

Assuming slow velocity regime, i.e., $v\ll c$, in a weak gravitational field, the space-time metric is approximated as
\begin{equation}
	g_{\mu\nu}=\eta_{\mu\nu}+h_{\mu\nu}
\end{equation}
where $\eta_{\mu\nu}$ is Minkowski metric and $h_{\mu\nu}$ is the corresponding metric perturbation. By imposing the harmonic gauge $\partial_{\mu} \overline{h}^{\mu\nu}=0$, the right-hand-side of the metric field equation can be linearized as follows
\begin{equation}
	G^{\mu \nu}+\Lambda g^{\mu\nu}\simeq -\frac{1}{2}\square \overline{h}^{~\mu\nu}+\Lambda \eta^{\mu\nu}
	\label{wfllambdafe}
\end{equation}
where
\begin{equation}
	\overline{h}^{\mu\nu}=h^{\mu\nu}-\frac{1}{2}\eta^{\mu\nu} h~~~~,~~~~h=h_{\alpha}^{\alpha}
\end{equation}
It is noteworthy that we have taken the Minkowski space-time as the background. Therefore, existence of $\Lambda$ should be treated as a perturbation. Consequently, we omit the term $\Lambda h^{\mu\nu}$ in the linearized field equation as it is a second order perturbation term. 

On the other hand, the perturbed metric can be written as
\begin{equation}
	ds^2=-\Big(1+\frac{2\Psi}{c^2}\Big) d(ct)^2+\Big(1-\frac{2\Phi}{c^2}\Big)\delta_{ij}dx^i dx^j
	\label{met} ,
\end{equation}
where $\Phi$ and $\Psi$ represent the gravitational potentials. In this case, the nonzero components of $\overline{h}^{\mu\nu}$ are
\begin{equation}
	\begin{split}
		& c^2 \overline{h}^{00}=-(\Psi+3\Phi)\\&
		c^2 \overline{h}^{ij}=(\Phi-\Psi)\delta^{ij}
	\end{split}
\end{equation}
Accordingly, the nonzero components of the Einstein tensor read
\begin{equation}
	\begin{split}
		c^2 G^{00}=\frac{1}{2}(\nabla^2 \Psi+3\nabla ^2 \Phi)
		\\c^2 G^{ij}=\frac{1}{2}(\nabla^2 \Psi-\nabla ^2 \Phi)\delta^{ij}
	\end{split}
\end{equation}
Note that the time derivatives of the potentials are small in the Newtonian limit. Therefore they can be safely neglected. Now, let's consider the energy-momentum tensor in the Newtonian limit. It is straightforward to verify that \citep{poisson2014gravity}
\begin{align*}
	\overset{(m)}{{T}^{00}} &=\rho c^2 +\mathcal{O}(1) \\
	\overset{(m)}{T^{0j}} &=\rho v^j c +\mathcal{O}(c^{-1}) \\
	\overset{(m)}{T^{ij}} &=\rho v^i v^j + p\, \delta ^{ij}  +\mathcal{O}(c^{-3}).
\end{align*}
{where $v^{j}=d x^j/dt$.}
By ignoring the terms $c^{-n}$ with $n\geq 3$, the $00$ component of the field equation \eqref{wfllambdafe} can be written as
\begin{equation}
	\frac{1}{2} (\nabla^2 \Psi + 3 \nabla^2 \Phi) = 8\pi G \rho + \Lambda c^2
\end{equation}
similarly the summation of $ii$ components takes the following form
\begin{equation}
	\frac{1}{2} (\nabla^2 \Psi - \nabla^2 \Phi) = -\Lambda c^2
\end{equation}
{Note that the $0j$ components do not contribute to the Newtonian limit}. By combining these two equations, the Poisson's equation can be obtained as follows
\begin{equation}
	\nabla^2\Psi = 4\pi G \rho - \Lambda c^2=4\pi G\Big(\rho+\rho_{\Lambda}+\frac{3p_{\Lambda}}{c^2}\Big)
	\label{lpe}
\end{equation}
where $\rho_{\Lambda}=\frac{\Lambda c^2}{8\pi G}$ is the dark energy density and $p_{\Lambda}=-c^2 \rho_{\Lambda}$. Of course, in galactic scales, the contribution of $\Lambda$ to Poisson's equation can be safely disregarded, as the dark energy density is significantly smaller than the matter density of the galaxies.

{Having reviewed the Newtonian limit of GR with the cosmological constant, let us now revisit the Newtonian limit of the CDE model. In this case, the metric field equation in the linear limit can be expressed as follows}

\begin{equation}
	-\frac{1}{2}\square \overline{h}^{~\mu\nu} = \frac{\kappa^2}{c^4} \Big(\overset{(b)}{T^{\mu\nu}} + \overset{(m)}{T^{\mu\nu}} + \overset{(\phi)}{T^{\mu\nu}}\Big).
	\label{feofCDE}
\end{equation}
We only need to linearize the energy-momentum tensor of the scalar field, as we have already computed the remaining terms. We assume that the scalar field comprises a cosmic background value that is a function of time, as well as a perturbation that is a function of both space and time
\begin{equation}
	\phi=\phi_0 (t) + \delta \phi(x^\mu)
\end{equation}
Accordingly, the energy-momentum tensor of the scalar field \eqref{emtsf} can be linearized as follows
\begin{align}\label{newt1}
	\overset{(\phi)}{T^{00}} &\simeq\frac{1}{2 c^2}\dot{\phi}_0^2+V_0 + V'_0 \,\delta\phi + \mathcal{O} (c^{-2})
	\\ \overset{(\phi)}{T^{ij}} &\simeq -\delta^{ij}\Big(-\frac{1}{2 c^2}\dot{\phi}_0^2+V_0 + V'_0 \delta \phi\Big) + \mathcal{O} (c^{-2})\label{newt2}
\end{align}
where $V_0=V(\phi_0)$, $V'_0=V'(\phi_0)$ and prime sign stands for derivative with respect to the scalar field. In this study, we disregard the time derivative of the scalar field perturbation due to its negligible changes in time compared to cosmic time. Consequently, by using the components of the energy-momentum tensor, namely equations \eqref{newt1} and \eqref{newt2}, the $00$ and sum of the $ii$ components of the field equation \eqref{feofCDE} can be written as
\begin{equation}
	\frac{1}{2} (\nabla^2 \Psi + 3 \nabla^2 \Phi) = 8\pi G (\rho_b + \rho_c) + \frac{8 \pi G}{c^2} \Big(\frac{1}{2c^2}\dot{\phi}_0^2+V_0 + V'_0 \,\delta\phi \Big)
\end{equation}
and 
\begin{equation}
	\frac{1}{2} (\nabla^2 \Psi - \nabla^2 \Phi) = \frac{8 \pi G}{c^2} \Big(\frac{1}{2 c^2}\dot{\phi}_0^2 - V_0 - V'_0 \delta \phi \Big)
\end{equation}
respectively. By combining these two equations, the generalized Poisson's equation can be derived in the following form
\begin{equation}
	\nabla ^2\Psi =4\pi G \Big( \rho _b +\rho_c +\rho _\phi +  \frac{3 p_{\phi}}{c^2} \Big) + \frac{8 \pi G}{c^2}V'(\phi_0)\delta\phi 
	\label{gpeq}.
\end{equation}
where the density $\rho_{\phi}$ and the pressure $p_{\phi}$ are defined as
\begin{equation}
	\rho_\phi = \frac{1}{2 c^4} \Dot{\phi}_0 ^2 + \frac{V(\phi_0)}{c^2}
\end{equation}
\begin{equation}
	p_{\phi} = \frac{1}{2 c^2} \Dot{\phi}_0 ^2 - V(\phi_0)
\end{equation}
The inclusion of the last terms in the right hand side of equation \eqref{gpeq}, despite their coefficient of $1/c^2$, is intended to enable a comparison with equation \eqref{lpe} and to facilitate a better understanding of the scalar field's role as dark energy. The dark energy contribution, $\rho _\phi + 3 p_{\phi}/c^2$, has the same role as the term  $\rho_{\Lambda}+\frac{3p_{\Lambda}}{c^2}$ in equation \eqref{lpe}. Similar to the case of $\Lambda$, we expect that this contribution is small compared to the matter density of the galaxy, namely $\rho _\phi + 3 p_{\phi}/c^2\ll\rho_b+\rho_c$. Likewise, the final term of equation \eqref{gpeq}, which incorporates a coefficient of $1/c^2$ and involves the small perturbation $\delta\phi$, can be safely ignored. Consequently, our generalized Poisson's equation keeps the Newtonian form
\begin{equation}
	\nabla ^2\Psi \simeq 4\pi G (\rho _b +\rho_c )
	\label{peq}
\end{equation}
However, the force exerted on each particle is not solely determined by $\nabla \Psi$. The correct form of the force is obtained from the Euler equation, which can be derived from the weak field limit of the conservation equation of the total energy-momentum tensor. By keeping the dominant terms in the conservation equation, which are of the order of $c$, while discarding those with a lower order, the $\mu=0$ component of equations \eqref{ceobm} and \eqref{ceodm} yields the continuity equation
\begin{equation}
	\frac{\partial \rho_A}{\partial t}+\nabla .( \rho_A\mathbf{v}_A)\simeq 0
	\label{ce}
\end{equation}
where $\rho_A$ can be either the baryonic matter density $\rho_b$ or the cold dark matter density $\rho_c$. Correspondingly, $\mathbf{v}_A$ can be the velocity of the baryonic fluid $\mathbf{v}_b$ or that of the dark matter fluid $\mathbf{v}_c$. Furthermore, by retaining the dominant terms (zeroth order of $c$) in three other components of the conservation equations, we obtain the modified Euler equation
\begin{equation}
	\rho_A \frac{\partial \mathbf{v}_A}{\partial t}+\rho_A (\mathbf{v}_A\cdot\nabla )\mathbf{v}_A=-\nabla p_A -\rho_A \nabla \Psi_A
	\label{mee}
\end{equation}
where the effective potentials are defined by
\begin{equation}
	\Psi_A=\Psi-\kappa \beta_A \delta\phi
	\label{effpsi}
\end{equation}
since $\beta_A$ is different for baryonic and dark matter fluids, these components experience different gravitational potential. The potential $\Psi$ is obtained from the Poisson equation \eqref{peq}. On the other hand, by disregarding the time derivative of $\delta \phi$ in equation \eqref{feq1}, we can derive the linearised version of the field equation for the scalar field
\begin{equation}
	\nabla ^2 \delta\phi =V''(\phi_0)\delta\phi-\kappa(\beta_b\rho_b +\beta_c\rho_c)
	\label{sfeomc2}
\end{equation}
This is the screened Poisson's equation for $\delta\phi$, and its solution can be expressed as
\begin{equation}
	\delta\phi=\frac{\kappa}{4\pi} \int \big[ \beta_b\rho_b(\mathbf{r}')+\beta_c\rho_c(\mathbf{r}')\big]\frac{e^{-\mid \mathbf{r}-\mathbf{r}'\mid/\lambda}}{\mid \mathbf{r}-\mathbf{r}'\mid} d^3 \mathbf{r}'
\end{equation}
where $1/\lambda$ is the screening parameter defined as 
\begin{equation}
	\frac{1}{\lambda}=\sqrt{V''(\phi_0)}
\end{equation}
Consequently, the potential $\Psi_A$  takes the following form
\begin{equation}
	\begin{split}
		\Psi_A(\mathbf{r})=& -G\int \frac{d^3 \mathbf{r}'}{\mid \mathbf{r}-\mathbf{r}'\mid}\Big[\rho_b(\mathbf{r}')+\rho_c(\mathbf{r}') \\& + 2\beta_A(\beta_b\rho_b(\mathbf{r}')+\beta_c\rho_c(\mathbf{r}'))e^{-\mid \mathbf{r}-\mathbf{r}'\mid/\lambda}\Big]
		\label{pore}
	\end{split}
\end{equation}
The first two terms represent the Newtonian potential, denoted as $\Psi_A^{N}$, while the last two terms can be regarded as the correction $\Delta \Psi_A$ arising from CDE. Thus, the total potential can be expressed as
\begin{equation}\label{pore2}
	\Psi_A(\mathbf{r})=\Psi_A^N(\mathbf{r})+\Delta \Psi_A(\mathbf{r})
\end{equation}
Finally, the gravitational field acting on the component $A$ can be obtained through the following relation
\begin{equation}
	\mathbf{g}_A=-\nabla\Psi_A
	\label{CDEforce}
\end{equation}
The attainment of the generalized potential relation in the CDE model provides a means to explore the dynamics of galaxies. The generalized potential relation can be employed to examine the impact of matter coupling with a scalar field on dynamical friction.

\section{Dynamical Friction in CDE}\label{DF}
Dynamical friction, as previously mentioned, plays a crucial role in the dynamics of galaxies, making its study necessary in various contexts. For instance, the motion of a star cluster within a galaxy, immersed in a background of other stars, is affected by the dynamical friction induced by the background particles. Similarly, if we consider a host galaxy's dark matter halo, the movement of satellite galaxies situated far from the halo's center is influenced by the dynamical friction caused by the halo. Furthermore, dynamical friction can also be assessed within a fluidic state. These conditions arise in the vicinity of galactic halos' centers, where massive black holes reside, and investigating dynamical friction through a hydrodynamic approach proves advantageous in these regions. In the context of this study, various papers have utilized hydrodynamic methods to examine different scenarios (see \cite{Berezhiani:2019pzd},  \cite{Lancaster:2019mde}, and \cite{Ostriker:1998fa}).

The calculation of dynamical friction is a complex matter that relies on multiple factors. Expressions for dynamical friction cannot be obtained for any arbitrary system. However, in the scenario where a point mass perturber is moving within a system of low mass particles or within a gaseous medium, it is feasible to determine the dynamical friction force by applying simplified assumptions. For instance, in the following subsection, we describe a simplified two-body scattering method that is employed to obtain the relationship of the dynamical friction force exerted on an object moving in the field of particles. This method neglects the self-gravity interaction between the background particles. On the other hand, as will be elucidated later, the analytical expression for the dynamical friction relationship utilizing the two-particle scattering approach exhibits inherent complexity. Consequently, in the second subsection, an alternative method, namely the gravitational wake method, which is comprehensively expounded upon in \cite{Tremaine:1984} and also employed in \cite{Roshan:2021ljs}, has been utilized to derive an expression for the dynamical friction relationship. Additionally, the dynamical friction can be calculated for gaseous systems by assuming that the background is a perfect fluid and the moving object is a point mass that disrupts the background, which is discussed in the third subsection.

\subsection{Chandrasekhar's approach: two-body problem in CDE}
In this section, we adopt the conventional approach outlined in \cite{Chandrasekhar:1943ys} to determine the dynamical friction force in CDE. The interaction between an object with a mass of $M$ and a velocity of $\mathbf{v}_M$, moving through a homogeneous population of background particles with a mass of $m$ and a velocity of $\mathbf{v}_m$, can be simplified by considering two-body encounters. By extending the results of the two-body scattering between $M$ and a background particle $m$ to encompass scattering from all particles, the dynamical friction force can be derived. Our aim is not to repeat the standard calculations of the friction force. Instead, we focus on the modifications that would arise in Chandrasekhar's formula due to the non-Newtonian potential \eqref{pore} experienced by the particles.

Referring to equation \eqref{pore}, the gravitational potential experienced by a baryonic point mass $M$ due to the presence of a point mass $m_A$ can be expressed as
\begin{equation}
	\Psi_{b}=-\frac{m_A G}{r}\Big(1+2 \beta_b \beta_A e^{- r/\lambda}\Big)
\end{equation}
Notice that $m_A$ can be a baryonic or dark matter point particle.
In this equation, $G$ represents the gravitational constant, $r$ denotes the distance between the two masses, and $\beta_b$ and $\beta_A$ are coupling parameters associated with the masses $M$ and $m_A$, respectively. It is easy to show that in the center of mass system the equation of motion of the reduced mass $\mu = \frac{m_A M}{m_A+ M}$ due to the gravitational force of the combined mass $m_A+M$, can be described by the following equation
\begin{equation}
	\ddot{\mathbf{r}}= -\frac{(m_A+M)G}{r^2}\Big[1+2\beta_b\beta_A  \Big(\frac{r}{\lambda}+1\Big)e^{-r/\lambda}\Big].
	\label{mf}
\end{equation}
When considering Newtonian gravity, or equivalently when $\lambda\rightarrow \infty$, we can integrate this equation of motion and determine the final change in velocity of the objects. However, the presence of an exponential term in the gravitational potential of the CDE model hinders the availability of an analytical solution. As a result, an approximate method must be employed to determine the dynamical friction force in CDE.

To address this issue, we can refer back to equation \eqref{mf} and consider the scenario where the gravitational constant $G$ is not constant but varies with distance. In this case, we follow the prescription introduced in \cite{Roshan:2021ljs}, and define the effective gravitational constant $G_{\text{eff}}$ as
\begin{equation}
	G_{\text{eff}}^A= [1+f_A(r)]G,
\end{equation}
where $f(r)$ is a function that accounts for the deviation from the Newtonian potential and is defined as follows
\begin{equation}
	f_A(r)= 2\beta_b\beta_A  \Big(\frac{r}{\lambda}+1\Big)e^{-r/\lambda}.
	\label{fa}
\end{equation}
To calculate the average value of the effective gravitational constant, note that $d<r<d_A$, where $d$ represents the size of object $M$ and $d_A$ represents the characteristic size of the system of particles with mass $m_A$. We take the average value of $f_A(r)$ within this interval to obtain the average value of the effective gravitational constant. The mathematical form of the standard Chandrasekhar formula for dynamical friction remains unchanged with this approximate approach. We only need to substitute $G$ with $\bar{G}_{\text{eff}}^A$, where $\bar{G}_{\text{eff}}^A=[1+\bar{f}_A(r)]G$ and $\bar{f}_A(r)$ is given by
\begin{equation}
	\begin{split}
		\bar{f}_A &=\frac{1}{d_A-d} \int_{d}^{d_A} f_A(r) dr  \\& = \frac{2 \beta_b \beta_A \left(e^{-\frac{d}{\lambda }} (2 \lambda +d)-e^{-\frac{d_A}{\lambda }} (2 \lambda +d_A)\right)}{d_A-d}
	\end{split}
\end{equation}
It is important to note that $\bar{f}_b>0$ irrespective of the sign of $\beta_b$, indicating that the effective gravitational constant experienced by the moving perturber $M$ due to the baryonic matter is always greater than $G$. As a result, CDE model increases the dynamical friction caused by the baryonic background particles. On the other hand, the sign of $\bar{f}_c$ depends on the sign of $\beta$ defined as 
\begin{equation}
	\beta= \beta_b \beta_c
\end{equation}
As a result, the dynamical friction induced by the dark matter background particles can be enhanced or weakened, depending on the sign of $\beta$. Now let's recall that the standard Chandrasekhar expression for dynamical friction is
\begin{equation}
	\mathbf{F}_{\text{DF}}= -4\pi G^2 M^2 \frac{\mathbf{v}_M}{v_M^3}\rho(<v_M) \ln \Lambda ,
\end{equation}
where $\rho(<v_M)$ denotes the density of the background particles with velocities smaller than the velocity of the perturber $v_M$
\begin{equation}\label{pert8}
	\rho(<v_M)= 4\pi \int_0 ^{v_M} v_m ^2 \mathcal{F}(v_m) dv_m
\end{equation}
$\mathcal{F}(v_m)$ is the distribution function of the background particles. On the other hand, $\ln\Lambda$ is the Coulomb logarithm given by
\begin{equation}
	\Lambda=\frac{ v_{\text{typ}}^2 D}{G(m+M)}
\end{equation}
where $v_{\text{typ}}$ is the typical velocity of the background particles with mass $m$, and $D$ is the characteristic size of the system. As already mentioned, to obtain the friction force in CDE we only need to $G\rightarrow G_{\text{eff}}^A$. Finally, keeping in mind that the galactic systems have both baryonic and dark matter components, the friction force in CDE takes the following form
\begin{equation}
\begin{split}
	\tilde{ \mathbf{F}}_{\text{DF}}=& -4\pi G^2 M^2 \frac{\mathbf{v}_M}{v_M^3}\Big((1+\bar{f}_b)^2\rho_b(<v_M) \ln \Lambda_b\\&+(1+\bar{f}_c)^2\rho_c(<v_M) \ln \Lambda_c \Big),
	\label{first_method}
\end{split}
\end{equation}
where
\begin{equation}
	\Lambda_b=\frac{ v_{b}^{2}\, d_b}{G(m_b+M)},\,\,\,\,\,\,  \Lambda_c=\frac{ v_{c}^{2}\, d_c}{G(m_c+M)}
\end{equation}
$v_c$ and $v_b$ are the typical velocity of the cold dark matter and baryonic matter respectively, and $d_c$ and $d_b$ are the corresponding characteristic sizes. Based on the observational bounds on the $\beta$ coupling coefficient discussed earlier, it is clear that $\bar{f}_b\propto \beta_b^2$ is small. Therefore, we can ignore this term from the dynamical friction force by approximating $(1+\bar{f}_b)^2\simeq 1$. On the other hand, in the self-gravitating systems studied in this paper, we expect the dark matter contribution to the dynamical friction to dominate the baryonic matter contribution. Thus, we can ignore the baryonic contribution. By doing so, we arrive at the following ratio between the dynamical friction force in CDE and Newtonian gravity for the same distribution of dark matter particles in both cases
\begin{equation}
	\mathcal{ R} =\frac{\tilde{F}_{\text{DF}}}{F_{\text{DF}}}=\Big(\frac{G^c _{\text{eff}}}{G}\Big)^2= 1 + 2 \overline{f}_c + \overline{f}_c^2  
\end{equation}
As already mentioned, depending on the sign of $\beta=\beta_b\beta_c$, this ratio can be greater or less than one. To be specific, $\mathcal{R}>1$ for $\beta>0$ and $\mathcal{R}<1$ for $\beta<0$.

\subsection{Gravitational wake}
In the previous section, we used an approximate method to derive the dynamical friction force. However, there is an alternative method that can provide a more precise result without the need for averaging over the effective gravitational constant. In this approach, we consider a subject body traversing through a field of stars characterized by a mass $m$. The objective is to determine the dynamical friction experienced by the subject body in this medium. To achieve this, we adopt the methodology outlined in the seminal work by  Tremaine and Weinberg \citep{Tremaine:1984}. This paper elucidates the dynamics of the background particles by investigating the changes in their momentum induced by the gravitational potential generated by the presence of a mass $M$. Given the fundamental principle of conservation of total momentum in a mechanical system, a comprehensive analysis of the momentum transfers among the particles allows us to quantify the changes experienced by the subject object. Consequently, it becomes imperative to calculate the perturbed gravitational potential that the background particles experience as a consequence of the presence of mass $M$.

Consider a test baryonic object $M$ that perturbs its surrounding. The potential felt by background particles with mass $m_A$ is given by $\Psi_A$, see equation \eqref{effpsi}. To proceed with our analysis, we work in the Fourier space, where we express any function $f(\mathbf{x},t)$ as
\begin{equation}\label{newf}
	f(\mathbf{x},t)=\frac{1}{(2\pi)^3}\int \hat{f}(\mathbf{k},t) e^{i  \mathbf{k}\cdot\mathbf{x}} d^3 k
\end{equation}
where $\hat{f}(k,t)$ is the Fourier integral transform of $f(\mathbf{x},t)$. Using the Fourier transform for equation \eqref{effpsi} we obtain
\begin{equation}
	\hat{\Psi}_{A}(\mathbf{k},t)=\hat{\Psi}(\mathbf{k},t)-\kappa \beta_A \hat{\delta\phi}(\mathbf{k},t).
\end{equation}
On the other hand, using the Fourier transform of equations \eqref{peq} and \eqref{sfeomc2} we find
\begin{equation}
	\begin{split}
		\hat{\Psi}(\mathbf{k},t) & =-\frac{\kappa^2}{2k^2}\Big(\hat{\rho}_b(\mathbf{k},t)+\hat{\rho}_c(\mathbf{k},t)\Big)
		\\ \hat{\delta\phi}(\mathbf{k},t) & =\frac{\kappa}{k^2+\lambda^{-2}}\Big(\beta_b \hat{\rho}_b (\mathbf{k},t)+\beta_c \hat{\rho}_c (\mathbf{k},t)\Big)
	\end{split}
\end{equation}
Consequently, the Fourier transform of the potential $\Psi_A$ is
\begin{equation}
\begin{split}
	\hat{\Psi}_{A}(\mathbf{k},t)=&-\frac{\kappa^2}{2k^2}\Big[\hat{\rho}_b(\mathbf{k},t) \Big( 1+2 \beta_A \beta_b \frac{k^2}{k^2+\lambda^{-2}}\Big)\\&+\hat{\rho}_c(\mathbf{k},t) \Big( 1+2 \beta_A \beta_c \frac{k^2}{k^2+\lambda^{-2}}\Big)\Big]
	\label{ftofeAp}
\end{split}
\end{equation}
The potential experienced by the background particle can be influenced by both other background particles and the test object. However, the impact of self-gravity is not taken into consideration in this study. Moreover, given that our test object is composed of baryonic matter, we specifically focus on the term in equation \eqref{ftofeAp} that incorporates the baryonic density. We consider a test object with a mass density given by $\rho_b = M \delta (\mathbf{x}-\mathbf{v}_{M} t)$. In this case, we have $\hat{\rho}_b= M e^{-\mathbf{k}\cdot\mathbf{v}_M t}$. Finally, using equations \eqref{ftofeAp} and \eqref{newf}, we can express the potential as
\begin{equation}
	\Psi_{A} (\mathbf{x},t) = - \frac{M G}{2 \pi^2} \int \frac{1+q_{A}(k)}{k^2} e^{i \mathbf{k}\cdot(\mathbf{x}- \mathbf{v}_{M} t)} d^3 k
	\label{ftofp}
\end{equation}
where $q_{A}(k)$ is defined as
\begin{equation}\label{qk}
	q_{A}(k)=2 \beta_A \beta_b \frac{k^2}{k^2+\lambda^{-2}}
\end{equation}
We neglect the interaction between background particles. Therefore the background particle $m_A$ moves on the path $\mathbf{x}=\mathbf{x}_0+ \mathbf{v}_{m_A} t$ with momentum $\mathbf{p}= m_A \mathbf{v}_{m_A}$ in the absence of the perturber $M$. Now, let us proceed with the calculation of the force exerted on $m_A$ due to the presence of the perturber $M$. This force is expressed as
\begin{equation}
	\frac{d\mathbf{p}}{dt} = -m_{A} \mathbf{\nabla} \Psi_{A}(\mathbf{x})
	\label{dfp}
\end{equation}
It is convenient to express the resulting deviation in the position and momentum of the particle $m_A$ in power series in terms of $G$ 
\begin{equation}
	\mathbf{x}=\mathbf{x}_0+ \mathbf{v}_{m_A} t+\sum_{i=1}^{n}\Delta_i \mathbf{x},~~~~~\mathbf{p}= m_A \mathbf{v}_{m_A}+\sum_{i=1}^{n}\Delta_i \mathbf{p}\label{pert1}
\end{equation}
where $\Delta_i\mathbf{x}$ and $\Delta_i\mathbf{p}$ are the corresponding perturbation of order $i$ in the position and momentum respectively. Substituting equation \eqref{pert1} into \eqref{dfp}, we obtain (for more details see \cite{Roshan:2021ljs})
\begin{equation}
	\frac{d (\Delta_1 \mathbf{p})}{dt} = i \frac{m_{A} M G}{2 \pi^2} \int \frac{1+q_{A}(k)}{k^2}  e^{i(\mathbf{k} \cdot \mathbf{x}_0 + \omega t)} \mathbf{k}\,d^3 k
	\label{pert2}
\end{equation}
where $\omega = \mathbf{k} \cdot (\mathbf{v}_{m_{A}} - \mathbf{v}_M)$. By integrating equation \eqref{pert2} twice over time, we obtain the first-order perturbation $\Delta_1 \mathbf{x}$ as follows
\begin{equation}\label{pert3}
	\Delta_1 \mathbf{x}= - i \frac{M G}{2 \pi^2} \int \frac{1+q_A(k)}{k^2 {\omega} ^2}  e^{i(\mathbf{k} \cdot \mathbf{x}_0 + \omega t)} \mathbf{k}\,d^3 k
\end{equation}
On the other hand, it is easy to show that the second-order perturbation in the momentum can be obtained from
\begin{equation}
	\frac{d (\Delta_2 \mathbf{p})}{dt} =- m_{A} ( \Delta_1 \mathbf{x} \cdot \nabla) \nabla \Psi_{A}(\mathbf{x}) 
	\label{pert21}
\end{equation}
We see that the tidal matrix $K_{ij}=\partial^2\Psi_A/\partial x^i \partial x^j$ appears in the right hand side and is computed over the unperturbed path $\mathbf{x}=\mathbf{x}_0+ \mathbf{v}_{m_A} t$. Using equation \eqref{ftofp}, the tidal matrix takes the following form
\begin{equation}
	K_{ij}=\frac{M G}{2 \pi^2} \int \frac{1+q_A(k')}{k'^2}  e^{i(\mathbf{k}' \cdot \mathbf{x}_0 + \omega' t)} k'_i k'_j \,d^3 k'\label{pert4}
\end{equation}
where $\omega' = \mathbf{k'} \cdot (\mathbf{v}_m - \mathbf{v}_M)$. It is useful to first replace $\mathbf{k}'$ by $-\mathbf{k}'$ in \eqref{pert4} and substitute it alongside with equation \eqref{pert3} into equation \eqref{pert21}. The resulting equation is 
\begin{equation}
	\begin{split}
		\frac{d (\Delta_2 \mathbf{p})}{dt}=& i m_{A} \Big(\frac{ M G}{2 \pi^2}\Big)^2
		\int \left[\frac{1+q_{A}(k)}{k^2}\right]  \left[\frac{1+q_A(k')}{k'^2 \omega^2}\right] \\& \times (\mathbf{k'} \cdot \mathbf{k}) e^{i[(\mathbf{k}-\mathbf{k'})\cdot \mathbf{x}_0 + (\omega-\omega')t]} \mathbf{k}'\, d^3 k d^3 k'
		\label{p2}
	\end{split}
\end{equation}
To find the net momentum change experienced by $M$ due to interaction with particles $m_A$ with velocity $\mathbf{v}_{m_A}$, we need to integrate $d(\Delta_1 \mathbf{p})/dt$ and $d(\Delta_2 \mathbf{p})/dt$ over all background particles and velocities. Let's first integrate over all the background particles with constant number density $n_A$
\begin{equation}
	\frac{d\mathbf{p}}{dt} = \int n_{A} \Big(\frac{d(\Delta_1 \mathbf{p})}{dt}+\frac{d(\Delta_2 \mathbf{p})}{dt}\Big)   d\mathbf{x}_0
\end{equation}
It is easy to show that the first term in the right hand side becomes zero. This can be shown by using the definition of the Dirac delta function
\begin{equation}
	\delta(\mathbf{k}-\mathbf{k'})=\frac{1}{(2\pi)^3}\int e^{i(\mathbf{k}-\mathbf{k'})\cdot\mathbf{x}_0}d\mathbf{x}_0\label{delta}
\end{equation}
This means that the DF force does not appear at the linear order of $G$. Therefore, let's proceed to calculate the second term. Using equations \eqref{p2} and \eqref{delta}, we find 
\begin{equation}
	\frac{d\mathbf{p}}{dt}  = \frac{2 i}{ \pi} m_{A}n_{A}  M^2 G^2 \int \frac{[1+q_{A}(k)]^2}{k^2 \omega^2} \mathbf{k} d^3 k
	\label{coi}
\end{equation}
According to \cite{Roshan:2021ljs}, the above equation can be rewritten as
\begin{equation}
	\frac{d \mathbf{p}}{dt}  = 4\pi m_{A}n_{A}  M^2 G^2 \frac{(\mathbf{v}_M-\mathbf{v}_{m_A})}{|\mathbf{v}_M-\mathbf{v}_{m_A}|^3}\int [1+q_{A}(k)]^2 \frac{dk}{k}
	\label{pert5}
\end{equation}
Now we integrate over all velocities $\mathbf{v}_{m_A}$. Assuming an isotropic velocity distribution for the background particles, the number density can be written as $n_A=\int f(v_{m_A}) d^3v_{m_A} $. Consequently, to account for all the velocities, we need to do the following replacement in \eqref{pert5}
\begin{equation}\label{pert6}
	n_A\frac{(\mathbf{v}_M-\mathbf{v}_{m_A})}{|\mathbf{v}_M-\mathbf{v}_{m_A}|^3}\mapsto \int f(v_{m_A}) \frac{(\mathbf{v}_M-\mathbf{v}_{m_A})}{|\mathbf{v}_M-\mathbf{v}_{m_A}|^3} d^3v_{m_A}
\end{equation}
This replacement allows us to integrate over all velocities and obtain the net momentum change experienced by $M$ due to interaction with all background particles. On the other hand, using the Newton's shell theorem, one may write
\begin{equation}\label{pert7}
	\begin{split}
		\int f(v_{m_A}) \frac{(\mathbf{v}_M-\mathbf{v}_{m_A})}{|\mathbf{v}_M-\mathbf{v}_{m_A}|^3}& d^3v_{m_A} =\\& 4\pi \frac{\mathbf{v}_M}{v_M^3}\int_0^{v_M}f(v_{m_A})v_{m_A}^2 d v_{m_A}
	\end{split}
\end{equation}
Applying \eqref{pert6} to \eqref{pert5} and using \eqref{pert8}, we find the dynamical friction force The resulting equation is 
\begin{equation}
	\tilde{ \mathbf{F}}_{\text{DF}}=-\frac{d\mathbf{p}}{dt}= -4 \pi M^2 G^2  \frac{\mathbf{v}_M}{v_M ^3} \rho_A(<v_M) I_A
	\label{ptot1}
\end{equation}
where $I_A$ is defined as
\begin{equation}
	I_A=\int_{k_{\text{min}}}^{k_{\text{max}}} \frac{(1+q_{A}(k))^2}{k} dk 
\end{equation}
Fortunately, it is straightforward to evaluate this integral. The result is
\begin{equation}
	I_A=\ln\Big(\frac{k_{\text{max}}}{k_{\text{min}}}\Big)+\mathcal{I}_A(k_{\text{max}})-\mathcal{I}_A(k_{\text{min}})
	\label{q1}
\end{equation}
where the function $\mathcal{I}_A$ is defined as
\begin{equation}
	\mathcal{I}_A(k)=2\beta_A \beta_b \left(\frac{\beta_A\beta_b}{1+k^2\lambda^2}+(1+\beta_A \beta_b)\ln (1+k^2\lambda^2)\right)
	\label{q2}
\end{equation}
As $\beta_A$ approaches zero, $\mathcal{I}_A$ becomes zero and equation \eqref{ptot1} simplifies to Chandrasekhar's expression with $\ln \Lambda=\ln (k_{\text{max}}/k_{\text{min}})$. It is important to note that the moving body $M$ moves within both baryonic and dark matter mediums. Therefore, the general form of the dynamical friction force takes the following form
\begin{equation}\label{gfdf}
	\tilde{ \mathbf{F}}_{\text{DF}} =-4\pi M^2 G^2\frac{\mathbf{v}_M}{v_M^3} \Big(\rho_b(<v_M) I_{b}+\rho_c(<v_M) I_{c} \Big)
\end{equation}
As previously discussed, the coupling between dark energy and baryonic matter is determined to be insignificantly small due to observational limitations. Consequently, terms involving the square or higher powers of the coupling parameter $\beta_b$ in of $I_b$ can be safely disregarded. With this approximation, our analysis will primarily focus on evaluating the contribution of dark matter coupling alone. Therefore, the ratio between dynamical friction force in CDE and Newtonian gravity takes the following form
\begin{equation}
	\begin{split}
		\mathcal{R}=&\frac{\tilde{ {F}}_{\text{DF}}}{F_{\text{DF}}}= 1+\frac{2\beta^2}{\ln{\frac{k_{\text{max}}}{k_{\text{min}} }} }\left(\frac{1}{1+k_{\text{max}}^2\lambda^2}-\frac{1}{1+k_{\text{min}}^2\lambda^2}\right)\\& +\frac{2\beta(1+\beta)}{\ln{\frac{k_{\text{max}}}{k_{\text{min}} }} }\ln{\left(\frac{1+k_{\text{max}}^2\lambda^2}{1+k_{\text{min}}^2\lambda^2}\right)} .
		\label{pert81}
	\end{split}
\end{equation}
where $k_{\text{min}}=2\pi /\max(G M/v_M ^2, d)$ and  $k_{\text{max}}=2\pi/ d_A$. In general, $d_A$, the maximum impact parameter, is size of the system. This implies that encounters with particles located anywhere within the system's boundaries can potentially contribute to the drag. The minimum impact parameter is determined by the maximum value between the $90^\circ$ deflection radius \citep{galacticdynamicsbook}, $G M/v_M ^2$, or the size of the object, denoted by $d$. Equation \eqref{pert81} turns out that this ratio is greater than one when $\beta > 0$. In contrast, when $\beta < 0$, the dynamical friction is weaker in CDE. This is entirely consistent with our findings in the previous subsection.

\subsection{Gaseous medium}
In this section we assume that the perturbing object moves within a perfect fluid with uniform density. The presence of a stationary object disturbs the density of the fluid in a symmetric manner. However, when the object moves through the background fluid at a certain speed, it disturbs the background in a heterogeneous manner. This heterogeneity is characterized by an increased density of the fluid in the form of a widening trail behind the moving object, commonly referred to as a wake. The presence of this dense wake leads to a decrease in the speed of the moving object, known as the effect of dynamical friction within the fluid. Consequently, this section aims to investigate disturbances in a fluid in order to calculate the dynamical friction. The significance of dynamical friction in gaseous systems, particularly those observed near galactic nuclei, serves as motivation for the calculations of this section.

We follow the same procedure implemented in \cite{Berezhiani:2019pzd} for the case of superfluids in order to derive the dynamical friction force in CDE. For this purpose, it is necessary to determine the dispersion relation that governs the propagation of disturbances in the environment. This requires a comprehensive Jeans analysis, which is applicable to an infinite uniformly distributed environment within the context of CDE. Using our results for the weak field limit of the theory, the basic hydrodynamic equations governing this system are formulated as follows
\begin{equation}
	\frac{\partial \rho_A}{\partial t}+\nabla .( \rho_A\mathbf{v}_A)\simeq 0
	\label{ce1}
\end{equation}
\begin{equation}
	\rho_A \frac{\partial \mathbf{v}_A}{\partial t}+\rho_A (\mathbf{v}_A\cdot\nabla )\mathbf{v}_A=-\nabla p_A -\rho_A \nabla \Psi_A
	\label{mee1}
\end{equation}
\begin{equation}
	\Psi_A=\Psi-\kappa \beta_A \delta\phi
	\label{effpsi1}
\end{equation}
where $\Psi$ and $\delta\phi$ are obtained from \eqref{peq} and \eqref{sfeomc2} respectively. These equations, along with the equation of state, depict the behavior of a perfect non-relativistic fluid within the framework of CDE. Let us consider a stationary background fluid that undergoes changes due to the presence of a perturbation. In the subsequent discussion, we denote the background functions with a subscript "0" and the perturbations with a subscript "1". Accordingly, we express the perturbed quantities as
\begin{equation}
	\begin{split}
		&  \rho_A=\rho_{A0} +\rho_{A1}
		\\& P_A=P_{A0}+P_{A1}
		\\& \mathbf{v}=\mathbf{v}_0+\mathbf{v}_1
		\\& \Psi=\Psi_0 +\Psi_1
		\\& {\phi}={\delta\phi_0}+\delta \phi_1
	\end{split}
\end{equation}
Now, the first-order perturbation equations can be readily derived as
\begin{equation}
	\frac{\partial \rho_{A1}}{\partial t}+\nabla \cdot (\rho_{A0} \mathbf{v}_1) +\nabla \cdot (\rho_{A1} \mathbf{v}_0) = 0
	\label{cpe}
\end{equation}
\begin{equation}        
	\frac{\partial \mathbf{v}_1}{\partial t} + (\mathbf{v}_0 \cdot \nabla) \mathbf{v}_1 +(\mathbf{v}_1 \cdot \nabla) \mathbf{v}_0 = -\nabla h_{A1} - \nabla \Psi_{A1} 
	\label{epe}
\end{equation}
where $h_{A1}=c_{A}^2 \frac{\rho_{A1}}{\rho_{A0}^2}$. Where $c_A$ stands for the speed of sound in the system $A$. Employing the Jeans swindle \citep{galacticdynamicsbook}, we impose the conditions $\rho_{A0} =$ constant and $\mathbf{v}_0 = 0$ for the background system. Subsequently, we integrate the time derivative of equation \eqref{cpe} with the divergence of equation \eqref{epe} to yield
\begin{equation} 
	\ddot{\alpha}_{A}-c_{A} ^2 \nabla^2 \alpha_{A} - \nabla ^2 \Psi_{A1}=0
	\label{kn1}
\end{equation}
where $\alpha_A=\frac{\rho_{A1}}{\rho_{A0}}$ . On the other had, it proves useful to transform to the Fourier space and utilize  
\begin{equation}
	\Psi_{A1} (\mathbf{x},t) = \frac{1}{(2\pi)^4} \int  \hat{\Psi}_{A1} e^{i (\mathbf{k}\cdot\mathbf{x}-\omega t)} d^3 k\, d\omega 
	\label{dis1}
\end{equation}
and 
\begin{equation}
	\alpha_{A}(\mathbf{x},t)= \frac{1}{(2\pi)^4} \int \hat{\alpha}_{A} e^{i (\mathbf{k}\cdot\mathbf{x}-\omega t)} d^3 k \, d\omega.
	\label{kn2}
\end{equation}
Given equation \eqref{ftofeAp} for $\hat{\Psi}_{A1}$ in the condition that there is no external object, and by substituting equations \eqref{dis1} and \eqref{kn2} into equation \eqref{kn1}, the dispersion relation is obtained as
\begin{equation}
	\omega^2 -k^2 c_{A} ^2  + 4 \pi G \rho_{A0} \Big(1+2\beta_{A}^2 \frac{k^2}{k^2+\lambda^{-2}}\Big) =0 .
	\label{alphaeqj}
\end{equation}
It is now straightforward to determine the Jeans wave number ($\tilde{k}_J$) in order to define the maximum stable wavelength ($\Tilde{\lambda}_J$). This can be achieved by setting the frequency parameter $\omega$ equal to zero and solving the dispersion relation for $\tilde{k}_J$.  Here, $\Tilde{\lambda}_J$ represents the Jeans wavelength in CDE framework. Therefore, the Jeans wave number can be obtained from the following relation
\begin{equation}
	\frac{\Tilde{k}_J^2}{k_J^2}= 1+ 2\beta_A^2 \frac{\Tilde{k}_J^2}{\Tilde{k}_J^2+\lambda^{-2}}
	\label{dis2}
\end{equation}
where $k_J^2=4\pi G \rho_{0A}/c_A^2$ represents the Newtonian Jeans wave number, and $\lambda_J=2\pi/k_J$ is the corresponding Newtonian wavelength.
Let us define the dimensionless parameters $\hat{k}$ and $\hat{\lambda}$ as $\hat{k}=(\Tilde{k}_J/k_J)^2$ and $\hat{\lambda}=(\lambda_J/2\pi\lambda)^2$. This way, the solution of \eqref{dis2} takes the following form 
\begin{equation}
	\hat{k}=\frac{1}{2}\Big(1+2\beta_A^2 -\hat{\lambda}+\sqrt{4\hat{\lambda}+(1+2\beta_A^2-\hat{\lambda})^2}\Big)>0
\end{equation}
As expected in the limit of $\beta_A\rightarrow 0$ or $\hat{\lambda}\rightarrow \infty$ (or equivalently $\lambda\rightarrow 0$) we recover the Newtonian case, namely $\hat{k}=1$. Notice that $\hat{k}>1$ for any choice of $\beta_A$ and $\hat{\lambda}$. Therefore the Jeans wavelength in CDE is always smaller than the Newtonian case. This directly means that the identical self-gravitating systems are more unstable in the context of CDE compared with the Newtonian case. 

Following the Jeans analysis within the CDE framework, we proceed to compute the perturbation induced in the background density by the presence of a moving object within it. This way, we can calculate the force acting on the object as a result of the disturbance created in the background. If we consider the background fluid density is homogeneous, $\rho_{A0}$, and we denote the density of the wake created as $\rho_{A1}$, and assuming that $v_M\neq 0$, the resulting force exerted by this wake on the moving object, leading to a decrease in its speed in direction of the motion, can be quantified as
\begin{equation}
	\tilde{F}_{DF}=\frac{M}{v_{M}} \dot{\Psi}_{M} (\mathbf{x},t)
	\label{eqdf}
\end{equation}
Here, $\dot{\Psi}_{M}$ represents the time derivative of the potential experienced by the object $M$ due to the wake, or alternatively $\rho_{A1}$, and is given by
\begin{equation}
	\dot{\Psi}_{M}(\mathbf{x},t)=-\frac{i}{(2\pi)^4}\int \omega \hat{\Psi}_{M}(\mathbf{k},\omega) e^{i(\mathbf{k}\cdot\mathbf{x}-\omega t)} d^3k~ d\omega
\end{equation}
The mass $M$ experiences an impact from the wake potential generated in the background fluid. Benefiting again from equation \eqref{ftofeAp}, the expression for $\hat{\Psi}_M$ can be written as
\begin{equation}
	\hat{\Psi}_{M}(\mathbf{k},\omega)=-\frac{\kappa^2 \rho_{A0}}{2k^2}  \Big( 1+2 \beta_A \beta_b \frac{k^2}{k^2+\lambda^{-2}}\Big)\hat{\alpha}_A(\mathbf{k},\omega)
\end{equation}
The derivation of a relation for $\hat{\alpha}_A$ becomes necessary in order to calculate dynamical friction. To achieve this,  we revisit equation \eqref{kn1} to incorporate the presence of the external object $M$ that moves with velocity $\mathbf{v}_M$. Therefore, the form of $\hat{\Psi}_{A1}$ changes due to the existence of $M$. According to equations \eqref{dis1} and \eqref{kn2} and by inserting them into equation \eqref{kn1}, the dispersion equation takes the following form
\begin{equation}
	\omega^2 \hat{\alpha}_A -k^2 c_{A} ^2 \hat{\alpha}_A -k^2 \hat{\Psi}_{A1} =0 .
	\label{alphaeqdf}
\end{equation}
Here $\hat{\Psi}_{A1}$ is Fourier transform of the perturbed potential that the background particles feel and it is generated by the moving baryonic body $M$ and the wake created in the background fluid itself. As a result, according to equation \eqref{ftofeAp}, in this case, $\hat{\Psi}_{A1}$ can be written as follows
\begin{equation}
	\begin{split}
		\hat{\Psi}_{A}(\mathbf{k},t)=&-\frac{\kappa^2}{2k^2}\Big[\hat{\rho}_M(\mathbf{k},t) \Big( 1+2 \beta_A \beta_b \frac{k^2}{k^2+\lambda^{-2}}\Big)\\& +\rho_{A0}\hat{\alpha}_A(\mathbf{k},t) \Big( 1+2 \beta_A \beta_c \frac{k^2}{k^2+\lambda^{-2}}\Big)\Big]
	\end{split}
\end{equation}
Inserting this equation into \eqref{alphaeqdf}, we obtain
\begin{equation}
	\begin{split}
		&\omega^2 \hat{\alpha}_{A}-k^2 c_{A} ^2 \hat{\alpha}_{A} + 4 \pi G \Bigg[\rho_{A0} \hat{\alpha}_A\Big(1+2\beta_{A}^2 \frac{k^2}{k^2+\lambda^{-2}}\Big)\\& +  \hat{\rho}_{M}\Big(1+2\beta_{b}\beta_{A} \frac{k^2}{k^2+\lambda^{-2}}\Big)\Bigg]=0
		\label{alphaeq}
	\end{split}
\end{equation}
Notice that for a point mass we have $\hat{\rho}_M(\mathbf{k},\omega)=\frac{M}{2\pi}\delta(\mathbf{k}\cdot\mathbf{v}_M-\omega)$. Thus, we can express $\hat{\alpha}_A(\mathbf{k},\omega)$ as
\begin{equation}
	\hat{\alpha}_{A}(\mathbf{k},\omega)=4 \pi G \Bigg[\frac{1+q_{A}(k)}{f^2(k) -\omega^2}\Bigg]\hat{\rho}_{M}(\mathbf{k},\omega)
\end{equation}
where 
\begin{equation}
	f^2(k)=c_{A}^2 k^2-4 \pi G \rho_{A0} \Big(1+2\beta_{A}^2 \frac{k^2}{k^2+\lambda^{-2}}\Big)
\end{equation}
Consequently, from equation \eqref{eqdf} the drag force due to the wake in each component of background is obtained as
\begin{equation}
	\begin{split}
		 \tilde{F}_{\text{DF}}=& i\frac{ 2 G^2 M^2 \rho_{A0}}{\pi v_M} \int  (1+q_{A}(k))^2 \frac{\omega e^{i(\mathbf{k}\cdot\mathbf{x}- \omega t)}}{k^2 (f^2(k)-\omega^2)} \\& \times \delta(\mathbf{k}\cdot\mathbf{v}_M-\omega) d^3k d\omega.
	\end{split}
\end{equation}
To solve the integral over $\omega$, we use the following contour integral
\begin{equation}
	\int \frac{\omega e^{-i\omega t} }{f^2(k)-\omega^2} \delta(\mathbf{k}\cdot\mathbf{v}_M-\omega) d\omega=\pi i e^{-i f(k) t}\, \delta(\mathbf{k}\cdot\mathbf{v}_M-f(k)) .
\end{equation}
\begin{figure*}[!]
	\center
	\includegraphics[scale=0.8]{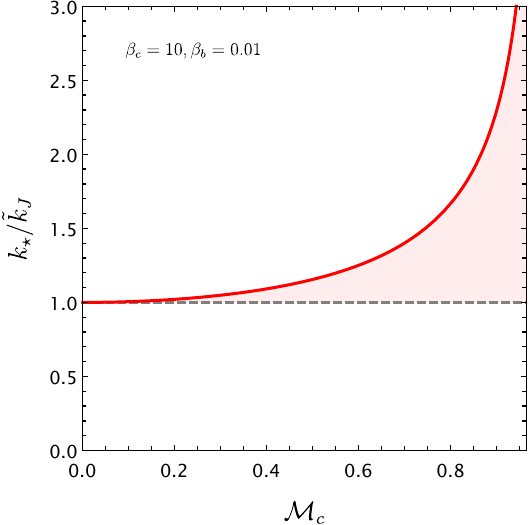}
	\hspace*{5mm}
	\includegraphics[scale=0.8]{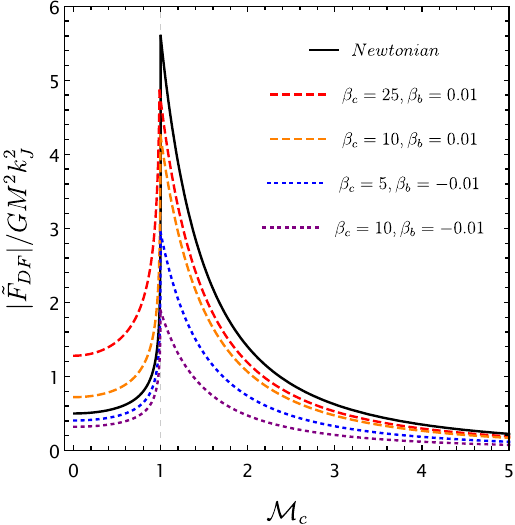}
	\caption{\textit{Left panel:} The ratio of $k_*$ to $\Tilde{k}_J$ in terms of $\mathcal{M}_c$. The shaded pink region delineates the permissible integration range for subsonic conditions. It is evident that with decreasing object velocity, the integration interval contracts, eventually approaching zero. \textit{Right panel:} The dynamical friction force in terms of $\mathcal{M}_c$. The dashed vertical line represents the critical value $\mathcal{M}_c=1$.The results do not change significantly with variations in $\lambda$, so we adopted a value of $\lambda=5.61\,$kpc.
	}
	\label{limitkandDfself}
\end{figure*}
Now, we can rewrite the expression for dynamical friction as follows

\begin{equation}
	\begin{split}
		 \Tilde{F}_{\text{DF}} =& -\frac{2 G^2 M^2 \rho_{A0}}{v_M} \int \frac{(1+q_{A}(k))^2}{k^2} \\& \times \delta(\mathbf{k}\cdot\mathbf{v}_M - f(k))  e^{i(\mathbf{k}\cdot\mathbf{x} -f(k) t)} d^3 k.
	\end{split}
\end{equation}
In order to simplify the analysis without loss of generality, let us consider the motion of the object in the $z$ direction and solve the integral using spherical coordinates. Therefore, we can write $\mathbf{x}= v_M t \hat{z}$ and $\delta(k v_M \cos{\theta} - f(k))= \frac{1}{k v_M} \delta\left(\cos{\theta} - \frac{f(k)}{k v_M}\right)$. We recall that our calculations in this section are based on the assumption that $v_M\neq 0$. Thus, we can rewrite the expression for $\Tilde{F}_{\text{DF}}$ as follows

\begin{equation}
	\begin{split}
		& \Tilde{F}_{\text{DF}}=-\frac{4\pi G^2 M^2 \rho_{A0}}{v_M^2} \int \frac{(1+q_{A}(k))^2}{k} \delta\left(\cos{\theta} - \frac{f(k)}{k v_M}\right) \\& e^{i(k v_M \cos{\theta} ~t-f(k) t)}dk ~d\cos\theta.
		\label{fdf4}
	\end{split}
\end{equation}
By simplifying this equation we obtain
\begin{equation}
	\Tilde{F}_{\text{DF}}=-\frac{4\pi G^2 M^2 \rho_{A0}}{v_M^2}\int_{k_{\text{min}}} ^{k_{\text{max}}} \frac{(1+q_{A}(k))^2}{k} dk
	\label{DFinGM}
\end{equation}
The analytic solution for this integral was previously derived in the preceding section, see equation \eqref{q1} and \eqref{q2}. The focus now shifts to the determination of the integration interval, which holds significance in this context. The range of wave number depends on the object and system characteristic size and the Jeans wavelength within the medium. However, a limitation arises when considering the self-gravity of the background system in perturbation analysis. This restriction emerges from the term $\delta\left(\cos{\theta}-\frac{f(k)}{k v_M}\right)$, in the calculation of the friction force. Notably, this integral imposes the following constraint 
\begin{equation}
	-1\leq\frac{f(k)}{k v_M}\leq1
\end{equation}
using this constraint we find
\begin{equation}
	k^2(1-\mathcal{M}_A^2)-k_J ^2 (1+Q(k)) \leq 0
	\label{re1}
\end{equation}
where $\mathcal{M}_A=\frac{v_M}{c_A}$ is the Mach number within the component $A$, and the dimensionless function $Q$ is defined as
\begin{equation}
	Q(k)=2\beta_{A}^2 \frac{k^2}{k^2+\lambda^{-2}}
\end{equation}
For a given scenario denoted as $\mathcal{M}_A>1$, the restriction \eqref{re1} is satisfied for all values of $k$, while for the case of $\mathcal{M}_A<1$, it is applicable solely to values less than $k_*$, where $k_*$ is the solution to the equation \eqref{re1}. In the absence of considering self-gravity we have $f(k)=c_A k$, and the condition \eqref{re1} leads to $\mathcal{M}_A>1$, indicating the absence of dynamical friction force on subsonic objects. Consequently, in light of these considerations, the allowable range of the integral is delineated as
\begin{equation}
	k_{\text{max}}= \min(\frac{2\pi}{d} , k_*) , ~~~~~~k_{\text{min}}=\max(\frac{2\pi}{d_A},\Tilde{k}_J)
\end{equation}

where, as already mentioned, $d_A$ is the characteristic size of the background system, and $d$ is the size of the object. Figure \ref{limitkandDfself} illustrates the influence of object velocity on the integration interval. Notably, as the speed decreases, the integration range contracts, leading to a corresponding decrease in the dynamical friction force on the object, as demonstrated in Figure \ref{limitkandDfself}. In contrast, as the speed increases, the integration interval expands until reaching a point where the value of $k_*$ equals $2\pi/d$. Beyond this juncture, the integration interval remains constant. These investigations highlight distinct behaviors in the final dynamical friction function for values of $\mathcal{M}$ less than 1 compared to those exceeding 1, as clearly delineated in the right panel of Figure \ref{limitkandDfself}. It is noteworthy that the plot representing the function at $\mathcal{M}$ equals 1 exhibits both a breaking in function and continuity.
We observe that when $\beta < 0$, the dynamical friction is less than that in the Newtonian case for both subsonic and supersonic speeds. Specifically, as $|\beta|$ increases, the dynamical friction decreases. Conversely,in this case, when $\beta > 0$, the friction is lower than the Newtonian case for supersonic speeds, while it is greater for subsonic speeds\footnote{In the supersonic case the dependence of dynamical friction on increasing $\beta$ exhibits object size-dependence. The friction force may increase or decrease as $\beta$ increases. Furthermore, even for small objects, dynamical friction can exceed its Newtonian counterpart at sufficiently large values of $\beta$. However, within the considered object size and $\beta$ ranges in this work, the magnitude of dynamical friction for supersonic case consistently remains below that predicted by Newtonian dynamics.}. This result is obtained under the condition that the system size exceeds the Jeans length. This leads to the dynamical friction described in equation \eqref{DFinGM} exhibiting a dependence on two factors: the integration interval and an additional term in equation \eqref{q1}. The $\beta$ parameter influences both of these factors. The combined effect of these two factors determines whether the dynamical friction is enhanced or suppressed relative to its Newtonian counterpart. For supersonic regimes, the CDE model predicts that the background system becomes increasingly unstable for $\beta\neq 0$. This results in a smaller integration interval, reducing the dynamical friction. On the other hand, as $\beta <0$, the additional term reduces, while for values $\beta>0$, it enhances. As illustrated in right panel of Figure \ref{limitkandDfself}, the effect of the integration interval dominates, resulting in a decrease in dynamical friction compared to the Newtonian case for $\beta <0$.  

The nonzero value of the friction at $v_M=0$ may seem puzzling. It should be noted that our calculation started from equation \eqref{eqdf} where we assumed $v_M\neq 0$. For the case of $v_M=0$, this equation should be written as $\tilde{F}_{DF}=-M \nabla \Psi_M$. In this case it is easy to shown that $\hat{\rho}_M(\mathbf{k},\omega)=\frac{M}{2\pi} \delta(\omega)$ leading to a spherically symmetric and static disturbance in the field system. As a result, the \eqref{fdf4} takes the following form
\begin{equation}
	\tilde{F}_{\text{DF}}=-4\pi G^2 M^2 \rho_{A0} \int k \cos{\theta} ~\delta(f(k))~e^{-if(k) t} dk ~d\cos\theta.
\end{equation}
As expected, the dynamical friction vanishes because of the integration over $\theta$.

It is important to note that if the system size is smaller than the Jeans length, then the integration interval remains constant, and the dynamical friction is solely determined by the additional term.  
In contrast, for subsonic regimes, the integration interval becomes independent of model parameters, eliminating its influence on the dynamical friction. Therefore, only the effect of the additional term persists, leading to an overall enhancement or suppression of dynamical friction based on the value of $\beta$.

As our final remark, returning to equation \eqref{DFinGM}, it is notable that the integral within this equation mirrors the integral discussed in the previous section. Assuming uniform mass for the moving object across all methodologies, the distinction in calculating the dynamical friction force between the current and previous section lies in the lower limit of the integral and the field density function utilized.

\section{Application to Fornax galaxy}\label{APP}
\begin{figure*}
	\centering
	\includegraphics[scale=0.7]{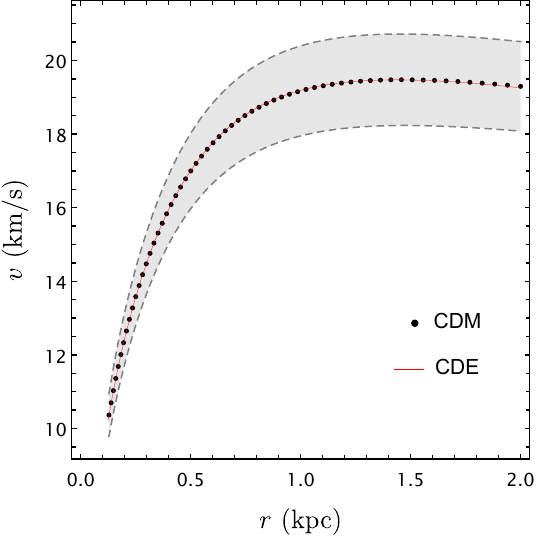}
	\includegraphics[scale=0.74]{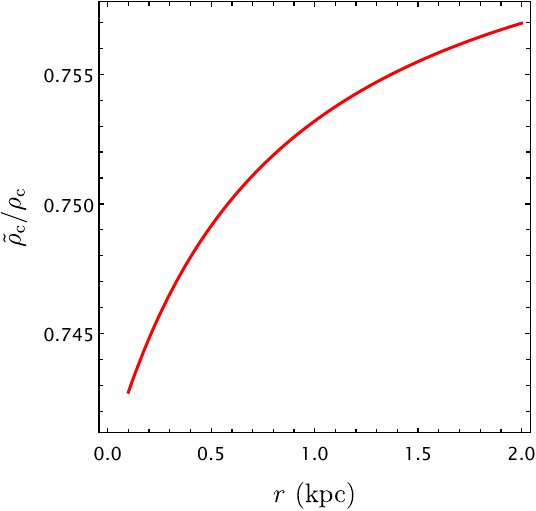}
	\caption{\textit{Left panel:} The panel illustrates the circular velocity data in the Newtonian regime, represented by black dots, alongside the velocity predicted by the CDE model, shown as a red curve. The density parameters within the CDE model were obtained through a fitting procedure, while the model parameters are fixed at $\beta=0.17$ and $\lambda = 5.61$ kpc. The shaded gray area indicates the error bounds related to the circular velocity data, which arise from uncertainties in the parameters $L$, $r_h$, and $V_{\text{max}}$. \textit{Right panel:} This panel displays the ratio of density in the CDE model, $\tilde{\rho}_c$, to the Newtonian density, $\rho_c$ as a function of radius.}
	\label{vfit}
\end{figure*}
We are now in a position to apply our findings to a real astrophysical system. The Fornax galaxy, a dwarf galaxy mainly composed of dark matter, serves as an excellent candidate for evaluating gravitational models. There has long been a puzzle regarding this galaxy: the globular clusters are situated at considerable distances from the center, even though dynamical friction from dark matter should have caused their orbits to decay toward the center. The aim of this section is to assess whether the CDE model effectively addresses this issue. Given that the dynamical friction relationship undergoes alteration in this model, it would be interesting to investigate whether the time taken by these clusters to reach the center differs from that in the CDM model. To achieve this, we require the characteristics, such as matter density, for both baryonic and dark matter. In the case of CDM, the dark matter density and associated free parameters that appear in the density profiles have been obtained using velocity dispersion measurements \citep{Walker:2009zp}. Conversely, the baryonic matter density, which is believed to follow a universal profile irrespective of the gravitational theory, has been obtained in \cite{Penarrubia:2007zx}.

However, in the context of CDE, the dark matter density is in principle different from CDM due to the alterations in the gravitational theory. Nevertheless, for practical purposes, we maintain the assumption that density profiles adhere to the same functional forms as in \cite{Walker:2009zp}, while allowing the parameters to remain flexible in order to align with observational data. We do not employ the Jeans equations and velocity dispersion observations to determine the free parameters of the dark matter density. Instead, we utilize the circular velocity $v^2=r \frac{d\Psi}{dr}$ obtained in CDM, which directly reflects the matter and ``gravity'' content of the galaxy, and find the free parameters by fitting the CDE model. The initial step involves obtaining the circular velocity relation in the CDE model. The circular velocity of the object $M$ is given by
\begin{equation}
	{v_M (r)}^2=r \frac{d\Psi_{M}(r)}{dr}.
	\label{rc}
\end{equation}
here, $\Psi_{M}(r)$ represents the gravitational potential exerted on the body $M$ situated at radius $r$ by the mass enclosed within the sphere $r$. This encompassed mass comprises both baryonic matter and dark matter. The baryonic stellar matter distribution in Fornax is modeled by the Plummer sphere \citep{Walker:2009zp}
\begin{equation}
	\rho_{b}(r)=\rho_{0}\left(1+\frac{r^2}{r_{h}^2} \right)^{-5/2} ,
	\label{density baryonic}
\end{equation}
where $\rho_0= \frac{3 L \Upsilon_*}{4\pi r_h^3}$. The total luminosity of Fornax is $L=(1.4\pm 0.4)\times 10^7 L_{\odot}$, the half-light radius is $r_h=668\pm 34$ pc, and the mass-to-light ratio is $\Upsilon_*\simeq 4.6 \Upsilon_{\odot}$ \citep{Penarrubia:2007zx}. Furthermore, the dark matter halo is modeled by NFW halo as follows
\begin{equation}
	\rho_{c}(r)=\rho_{s}\left( \frac{r_s}{r}\right)\left(1+\frac{r}{r_{s}} \right)^{-2} ,
	\label{density darkmatter}
\end{equation}
where $\rho_s=0.368 \frac{V_{\text{max}}^2}{G r_s^2}$ in which $r_s$ is the halo length-scale and $V_{\text{max}}$ is the maximum circular velocity associated with this halo. These parameters for the Fornax galaxy within the CDM model are $V_{\text{max}}=18\pm 1 \,\text{km\, s}^{-1}$ and $r_s=795\,$ pc \citep{Walker:2009zp}. As already mentioned, these parameters will be different in the context of CDE model. Let's denote them with tilde symbol as $\tilde{r}_s$ and $\tilde{V}_{\text{max}}$ or equivalently $\tilde{\rho}_s=0.368 \frac{\tilde{V}_{\text{max}}^2}{G \tilde{r}_s^2}$. By incorporating the density profiles of the two components into equations \eqref{pore} and \eqref{pore2}, a relationship for $\Psi_{M}(r)$ is derived. As mentioned earlier, the term containing $\beta_b ^2$ can be neglected in comparison to the other term in $\Delta \Psi_M$. It is easy to show that
\begin{equation}
	\Psi_M (r)=\Psi_{M}^N (r)+\Delta\Psi_{M} (r)
\end{equation}
where 
\begin{equation}\label{psiM}
	\Psi_{M}^N(x)=\frac{G L \Upsilon_*}{ \lambda(x_{h}^2+x^2)^{1/2}}+0.368 \frac{4 \pi \Tilde{V}^2_{\text{max}} \Tilde{x}_s}{x} \ln{(1+\frac{x}{\Tilde{x}_s})}
\end{equation} 
and 
\begin{equation}
	\Delta \Psi_{M}(x)=0.368 \frac{4 \pi \Tilde{V}^2_{\text{max}} \Tilde{x}_s  \beta}{x} \chi(x)
\end{equation}
where the function $\chi(x)$ is defined as
\begin{equation}
	\chi(x)= E(\Tilde{x}_s+x)+ E(-\Tilde{x}_s-x)-  e^{-x}E(-\Tilde{x}_s)- e^{-x}E(\Tilde{x}_s)
\end{equation}
and the dimensionless quantities $\tilde{x}_s$ and $x$ are defined as $\Tilde{x}_s=\Tilde{r}_s/\lambda$ and $x=r/\lambda$, and $E(X)$ is the product of the exponential function $e^{-X}$ and the exponential integral function $Ei(X)=-\int_{-X} ^{\infty} \frac{e^{-t}}{t} dt$, namely
\begin{equation*}
	E(X)=-e^{-X} \int_{-X} ^{\infty} \frac{e^{-t}}{t} dt .
\end{equation*}
Now using equation \eqref{rc}, we obtain the total circular velocity as follows
\begin{equation}
	v_{M}(x)^2= v_N (x)^2  +\Delta v (x)^2
	\label{vmm}
\end{equation}
Using the potential \eqref{psiM}, the Newtonian velocity $v_{N}$ is obtained
\begin{equation}
	\begin{split}
		v_{N}^2(x)=& \frac{G L \Upsilon_* x^2}{ \lambda(x_{\text{h}}^2+x^2)^{3/2}}+0.368\frac{4 \pi \Tilde{x}_s \Tilde{V}^2_{
				\text{max}}}{x} \\& \times \left( \ln{(1+\frac{x}{\Tilde{x}_s})}-\frac{x}{(\Tilde{x}_s+x)} \right), 
		\label{vnn}
	\end{split}
\end{equation}
On the other hand, the correction term in the circular velocity is given by
\begin{equation}
	\begin{split}
		\Delta v^2(x)=& 0.368\frac{4 \pi \Tilde{x}_s \Tilde{V}^2_{\text{max}} \beta}{x} \\& \times \Bigg(\frac{2 x}{x+\Tilde{x}_s} - (x+1)\chi(x)+2 x  E(-\Tilde{x}_s-x)\Bigg)
	\end{split}
\end{equation}
To obtain reference data for fitting the parameters of the CDE model, the circular velocity data for the galaxy were calculated using equation \eqref{vnn} with Newtonian parameters $r_s$ and $V_{\text{max}}$. We reiterate that the parameters related to the baryonic matter, i.e., $\rho_0$ and $r_h$, are identical in both the Newtonian and CDE model, as taken from \cite{Penarrubia:2007zx}. The CDE model (equation \eqref{vmm}) was then fitted to this reference data, aiming to achieve a close match between the velocities predicted by both models. This fitting process involved optimizing the dark matter density parameters, specifically focusing on the parameters denoted as $\Tilde{V}_{\text{max}}$ and $\tilde{r}_s$. Notably, the model's parameters, $\beta$ and $\lambda$, are set to $0.17$ and $5.61\,$kpc, respectively, as determined in \cite{deAlmeida:2018kwq}, which utilized the rotation curve data of spiral galaxies.

Through non-linear fitting in \texttt{Mathematica}, we derive suitable NFW density parameters as $\frac{\Tilde{V}_{\text{max}}}{V_{\text{max}}}=0.87\pm 1.7\times 10^{-4}$ and $\frac{\Tilde{r}_s}{r_s}=1.016\pm 8.0\times 10^{-4}$. The left panel in Figure \ref{vfit} illustrates the circular velocity data. The black dots represent the reference rotation curve, while the gray region specifies the corresponding errors. In contrast, the red curve depicts the fitted rotation curve in the CDE model. The right panel illustrates the ratio of dark matter density in CDE compared to Newtonian gravity. Interestingly, less dark matter is required in the CDE model.

\begin{figure*}[!]
	\centering
	\includegraphics[scale=0.75]{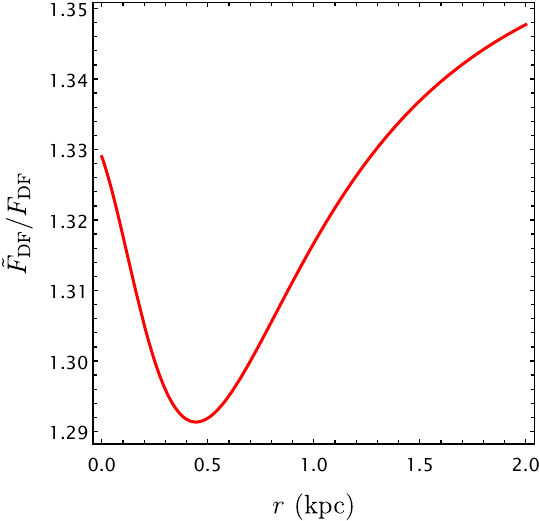}
	\hspace*{5mm}
	\includegraphics[scale=0.73]{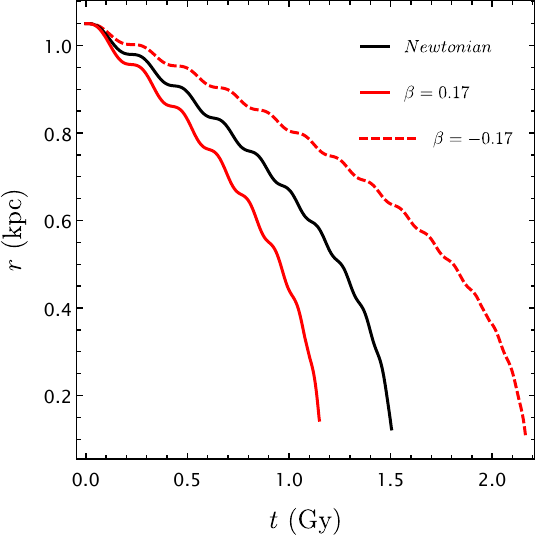}
	\caption{\textit{Left panel:} The left panel shows the ratio of the dynamical friction force in the CDE model to that in Newtonian gravity for the Fornax galaxy as a function of radius. \textit{Right panel:} This panel illustrates the time decay of one of the globular clusters in Fornax. The mass of this cluster is $1.82 \times 10^5 M_{\odot}$, and it is initially located at a distance of $r_0 = 1.05$ kpc from the center of the galaxy.}
	\label{cluster3time decay}
\end{figure*}

Now, let us assume that the expression for the dynamical friction force derived for a uniform background can be applied to the Fornax galaxy, where the density is isotropic but not homogeneous. While this assumption is not strictly accurate, it often provides a suitable qualitative analysis and insight of the significance of dynamical friction in the system \citep{galacticdynamicsbook}. To compute the time required for a globular cluster to move toward the galaxy's center, we first compare the magnitudes of the dynamical friction force in the CDE model and in Newtonian gravity. The results are presented in the left panel of Figure \ref{cluster3time decay}, where the ratio $\tilde{F}_{\text{DF}}/F_{\text{DF}}$ is plotted as a function of the distance from the center, $r$. Interestingly, although CDE requires less dark matter compared to CDM for this specific galaxy, the friction force is greater than the CDM model. This fact is attributed to the modifications that CDE induces to the gravitational law.

It is now evident that a globular cluster requires less time to reach the center in the CDE model. To illustrate this more explicitly, we solve the equations of motion for a test body with mass $M$ initially on a circular orbit with radius $r_0$. The particle experiences both the gravitational force \eqref{CDEforce} and the dynamical friction force \eqref{gfdf}. The right panel of Figure \ref{cluster3time decay} shows the temporal evolution of the distance from the galactic center for a globular cluster with a mass of $M = 1.82 \times 10^5\, M_{\odot}$ and an initial distance of $r_0 = 1.05$ kpc. These parameters closely resemble the current characteristics of one of the globular clusters in the Fornax galaxy. Compared to the Newtonian model, the CDE model exhibits a more pronounced rate of angular momentum loss due to higher dynamical friction for the globular cluster. This leads to an accelerated inward migration of the cluster towards the galactic center. Evidently, the CDE model demonstrates a decrease in the time required for the cluster to reach the center, thereby exacerbating the issue within the context of the Fornax galaxy.

\begin{figure*}
	\centering
	\includegraphics[scale=0.5]{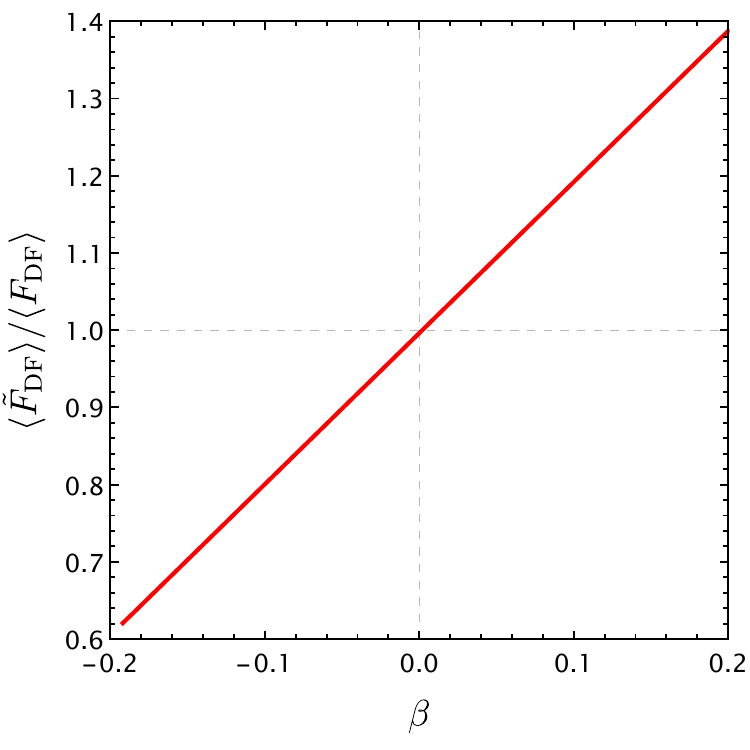}
	\hspace*{5mm}
	\includegraphics[scale=0.5]{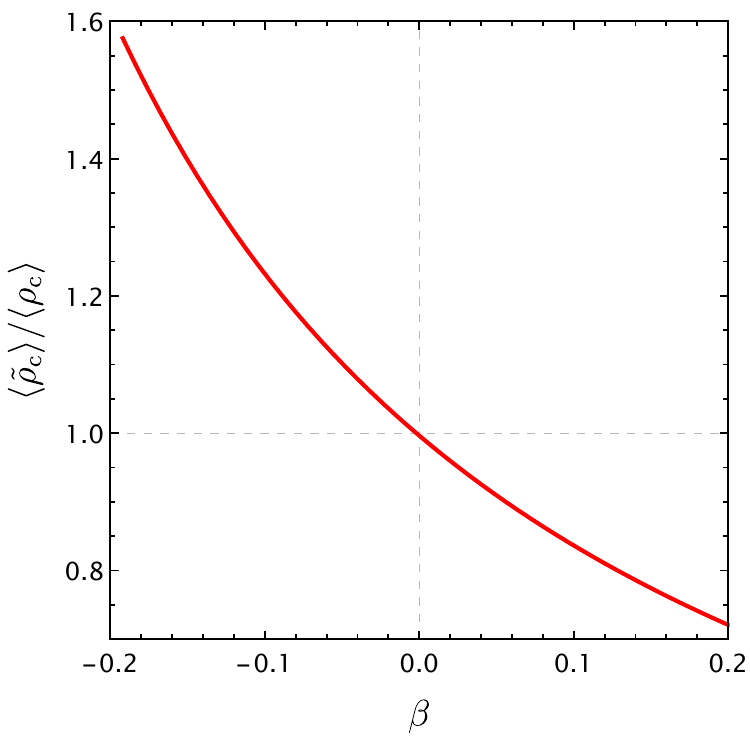}
	\caption{\textit{Left panel:} The average value of the ratio of dynamical friction in the CDE model to that in Newtonian gravity for the Fornax galaxy as a function of $\beta$. \textit{Right panel:} The corresponding average value of dark matter density as a function of $\beta$.} 
	\label{mean}
\end{figure*}

It appears that the magnitude of the dynamical friction force in CDE is more sensitive to the value of $\beta$ rather than $\lambda$. It is informative to consider a case where $\lambda$ is fixed at $\lambda = 5.61$ kpc and the $\beta$ parameter is varied. It is important to note that this parameter, in principle, can also take negative values. We repeat our analysis for the Fornax galaxy, this time implementing different values of $\beta$. For each value of $\beta$, we determine the Fornax dark matter density in the CDE model and the corresponding dynamical friction force. We then calculate the radial mean values of $\tilde{\rho}_c$ and the dynamical friction force $\tilde{F}_{\text{DF}}$ in the radial interval $r \in [0, 2\, \text{kpc}]$. The results are shown in Figure \ref{mean}. The right panel clearly shows that $\beta > 0$ yields less dark matter in Fornax compared to CDM. Conversely, $\beta < 0$ leads to more dark matter compared to CDM. However, interestingly, $\beta > 0$ enhances the dynamical friction, while $\beta < 0$ reduces it within Fornax, as seen in the left panel of Figure \ref{mean}. This implies that to resolve the Fornax puzzle, the necessary condition is to consider negative values of $\beta$. Of course, this choice will impact the rotation curve data, where $\beta > 0$ is preferred.

As a final remark in this section, let us consider how the cosmological constraints on the CDE coupling parameters change the dynamical friction within Fornax.
Cosmological constraints on $\beta_c$ have been outlined in \cite{Gomez-Valent:2020mqn,Pettorino:2013oxa,Xia:2013nua,Planck:2015bue}. We note that the $\beta_b$ parameter is well constrained by local gravitational experiments \citep{Amendola:1999er}. The cosmological constraints on $\beta_c$ have been derived using both exponential and power-law potential functions. Although different limits for $\beta_c$ have been obtained in different studies, here, we will refer to one of the most recent and restrictive limits reported in \cite{Gomez-Valent:2020mqn} that analyses incorporate CMB data from Planck 2018, as well as observations from the SH0ES \footnote{SNe, $H0$, for the Equation of State of dark energy} and H0LiCOW \footnote{$H0$ Lenses in COSMOGRAIL’s Wellspring} collaborations. The analysis reveals a value of $\beta_c = 0.010^{+0.003}_{-0.009}$ when considering CMB lensing and $\beta_c = 0.015^{+0.007}_{-0.008}$ when excluding it from the model.
Given these limitations, we set the value of $\beta_c$ to $0.02$ and observe that the CDE model exhibits minimal deviation from the Newtonian physics. Consequently, if the cosmological constraints are assumed for $\beta_c$, the effects of CDE are expected to be negligible at the galactic level.

\section{Conclusions}

We have calculated the dynamical friction force within the framework of the CDE model. Initially, we derived the Newtonian limit of the model. Subsequently, we calculated the gravitational potential experienced by each component, baryonic and dark matter, within a galaxy. The central ingredient of the model is that both baryonic matter and dark matter interact with dark energy.

Naturally, the modified gravitational potential in the CDE model implies a corresponding modification to the standard dynamical friction force. To investigate this, we adopted three approaches:
i) We examined the two-body problem in CDE and followed an approximate method to derive the dynamical friction force, as shown in equation \eqref{first_method}.
ii) In the second approach, we relaxed the restrictive assumptions of the first method and generalized the approach presented in \cite{Tremaine:1984}. This allowed us to derive an analytical expression for dynamical friction, as given in equation \eqref{ptot1}.
iii) Focusing on fluid systems, we applied perturbative Jeans analysis in CDE to compute dynamical friction. The mathematical form of the dynamical friction force obtained through this method is consistent with that of the second method, as shown in equation \eqref{DFinGM}. However, there are specific differences, particularly in the limits of the underlying integral.

The first two methods, applied to a particle system, demonstrate that when comparing the same background system in Newtonian gravity and CDE, the dynamical friction is stronger in CDE if $\beta>0$. Conversely, the dynamical friction force is weaker in CDE if $\beta<0$. However, the situation in fluid systems differs. Our analysis using the third approach reveals that when $\beta>0$, the dynamical friction force is stronger in CDE only for subsonic velocities of the perturber $M$. For $\beta<0$, the force is weaker in CDE at subsonic speeds. In the case of supersonic velocities, both $\beta<0$ and $\beta>0$ result in weaker friction in CDE compared to the Newtonian case.

Finally, we applied our findings to the long-standing puzzle of globular cluster positions in the Fornax dwarf galaxy, examining whether CDE alleviates this issue. Using observed velocity dispersions, the baryonic and dark matter distributions of this galaxy are well-established in the standard cold dark matter model. Utilizing these density profiles, we derived the circular velocity within the galaxy. As a final step, we fitted the CDE model, which incorporates both a dark matter halo and corrections to the gravitational force, to the rotational velocity. This approach allowed us to determine the dark matter density in CDE and calculate the dynamical friction within Fornax in the CDE framework. Considering the observational constraints on $\beta>0$ and $\lambda$ derived from galactic rotation curves, we find that although less dark matter content is expected in Fornax, the dynamical friction is stronger compared to the Newtonian case. Consequently, CDE reduces the orbital decay timescale. Despite being inconsistent with rotation curve data, we also examined the case where $\beta<0$. In this scenario, the dark matter density is higher in CDE compared to cold dark matter, while the friction is weaker. This results in an increased orbital decay timescale. It's worth noting that our results were significantly more sensitive to changes in $\beta$ than in $\lambda$. We also demonstrated that if the cosmological constraints on the $\beta_c$ parameter are assumed, it becomes impossible to distinguish between the CDE model and the standard model using the dynamical friction force in Fornax.

\section*{Acknowledgements}
This work is supported by Ferdowsi University of Mashhad under Grant No. 58497 (28/08/1401).

\bibliography{refs}

\end{document}